# THE EFFECT OF GEOMETRY ON CONCENTRATION POLARIZATION IN REALISTIC HETEROGENEOUS PERMSELECTIVE SYSTEMS

Yoav Green, Shahar Shloush, and Gilad Yossifon*

Faculty of Mechanical Engineering, Micro- and Nanofluidics Laboratory, Technion–Israel Institute of Technology, Technion City 32000, Israel

This study extends previous analytical solutions of concentration-polarization occurring solely in the depleted region, to the more realistic geometry consisting of a three dimensional (3D) heterogeneous ion-permselective medium connecting two opposite microchambers (i.e. 3 layers system). Under the local electro-neutrality approximation, the separation of variable methods is used to derive an analytical solution of the electro-diffusive problem for the two opposing asymmetric microchambers. Assuming an ideal permselective medium allows for the analytic calculation of the 3D concentration and electric potential distributions as well as a current-voltage relation. It is shown that any asymmetry in the microchamber geometries will result in current rectification. Moreover, it is demonstrated that for non-negligible microchamber resistances the conductance does not exhibit the expected saturation at low concentrations but instead shows a continuous decrease. The results are intended to facilitate a more direct comparison between theory and experiments as now the voltage drop is across a realistic 3D and 3-layer system.

## I. Introduction

The passage of an electric current through a permselective medium (membranes or nanochannels) under an applied electric field, is characterized by the formation of ionic concentration gradients which result in regions of depleted and enriched ionic concentration at opposite ends of the medium [1]. The formation of these concentration gradients and resulting electric current are collectively termed concentration polarization (CP). In the low-voltage region, the current-voltage (I-V) behavior is approximately Ohmic until the diffusion limited current saturates when both ion concentrations are completely depleted at the surface [2]. Since in the current study we aim at developing an analytical description of the CP phenomenon for realistic 3D and 3-layers systems (i.e. permselective medium connected by two opposing microchambers), we focus on the underliming response of the system where the effects of space charge layer (SCL) [3] and electro-convection [4–9] can be safely ignored. Thus, justifying the use of the local electroneutrality (LEN) approximation.

In heterogeneous permselective systems, i.e. membranes with a partly conducting surface area [10] or fabricated micro-nanochannel systems [11–15], the electroconvective mechanism induces strong corner vortices that stir the flow and in turn control the length of the diffusion layer. However, aside from these electro-convection effects, field focusing effects alone, stemming from the heterogeneity of 2D [16–18] and 3D [19] geometries, can affect the electro-diffusive solution to yield CP with much larger concentration gradients. These significantly larger gradients lead to a corresponding increase of the current density and result in an effectively shorter diffusion layer (DL) length, thereby reducing the importance of the electro-convective contribution.

Asymmetric microchambers geometries [20] and asymmetric micro-nanochannel interfaces [14] have experimentally been shown to rectify the current. It is expected that due to the existence of CP, the electro-diffusive problem, solved herein, will be able to capture current rectification. Although, electro-convection effects may enhance current rectification, these usually become significant only at sufficiently high voltages when the SCL appears. Other previously studied symmetry breaking conditions that rectify current include asymmetric concentrations in opposing reservoirs [21], modulation of the nanochannel surface charge [22], non-straight (commonly conical or funnel-shaped) nanochannel/nanopore geometry [23,24].

The LEN approximation to the electro-diffusive problem is solved analytically in an extended 3D heterogeneous geometry that includes three layers (both the anodic and cathodic microchambers that are connected via a permselective medium - see FIG. 1). Solving for a 3 layer system versus the commonly solved 1 layer system, consisting solely of the depleted layer [3,11,19,25,26], will better facilitate the comparison of theoretical models to experiments [5,6,12,13,27–33] and simulations [34–37].

A recently published paper [38] (published after the submission of this work) has solved a similar problem of a three-layered system in two-dimensions (2D). Our study is hence more general in terms of geometry (3D), while the former [38] is more general in terms of the counterion transport number as they account for a non-ideal membrane permselectivity. In addition, the focus of these works is substantially different. Their work [38] is focused on the variation of the system permselectivity in the course of CP, and hence, mainly described its effect on the counterion transport number. The current study focuses on the effect of a more realistic micro-permselective medium geometry (in 2D both the microchamber length and height vary, while in 3D also their widths) on the current-voltage (I-V) response. In particular, we study the effect of increased heterogeneity (i.e. field focusing) on the current density, current rectification due to asymmetric microchamber geometry, and an interesting break of conductivity saturation at low concentrations.

In Section II we will define the theoretical model and present its solution. In Section III we will provide details on numerical simulations conducted for verification of our theoretical model. In Section IV we shall present the verification of our model as well as additional key results. Concluding remarks will be given in Section V.

## II. Theoretical model

### A. Assumptions and governing equations

The steady state electrokinetic ionic transport of a symmetric and binary electrolyte $(z_+ = -z_- = 1)$, with ions of equal diffusivity $(D_\pm = D)$, is governed by the dimensionless Poisson-Nernst-Planck (PNP) equations

$$\nabla \cdot [\nabla c_+ + c_+ \nabla \phi] = -\nabla \cdot j_+ = 0, \quad (1)$$

$$\nabla \cdot [\nabla c_- - c_- \nabla \phi] = -\nabla \cdot j_- = 0, \quad (2)$$

$$\nabla^2 \phi = -\frac{\rho_e}{2\delta^2}, \quad (3)$$

wherein Eqs.(1) and (2) are the Nernst-Planck equations satisfying the continuity of ionic fluxes under steady-state conditions. The cationic and anionic concentrations, $\tilde{c}_+$ and $\tilde{c}_-$, respectively, have been normalized by the bulk concentration $c_0$. The spatial coordinates have been normalized by the DL length $\tilde{L}$, while the ionic fluxes have been normalized by $Dc_0 / \tilde{L}$. The tilde stands for the parameter in its dimensional form. Equation (3) is the Poisson equation for the electric potential, $\tilde{\phi}$, which has been normalized by the thermal potential $RT/F$ where $R$ is the universal gas constant, $T$ is the absolute temperature and $F$ is the Faraday constant. The charge density, $\rho_e = c_+ - c_-$, appearing in Eq. (3) is normalized by $zFc_0$. The normalized Debye layer is $\delta = \lambda_D / \tilde{L}$, with $\lambda_D = \sqrt{\varepsilon_0 \varepsilon_r RT / 2F^2 c_0}$ where $\varepsilon_0$ and $\varepsilon_r$ are the permittivity of vacuum and the relative permittivity of the electrolyte, respectively.

Using the LEN approximation [2,3,39,40], we presume that $\delta \ll 1$ (or alternately $\delta^2 \nabla^2 \phi \sim 0$) within the microchambers, simplifies the equations by replacing the Poisson equation (Eq. (3)) with the approximate condition $c_+ = c_- = c$. Thus, Eqs (1)-(3), assuming ideal cation permselectivity, $j_- = 0$, reduce to

$$\nabla^2 c = 0, \quad (4)$$

$$\phi = \ln c + \bar{\phi}, \qquad (5)$$

where the potential is defined up to an integration constant $\bar{\phi}$ and $c$ is the microchamber concentration. Eqs. (4) and (5) are correct for the microchambers where the concentration is allowed to vary. This is in contrast to the ideal permselective nanoslot, wherein the concentration does not vary. This point will be expanded upon in Section II.C.

## B. Geometry and boundary conditions

Our model consists of a 3-layers system in which two microchambers are connected by a straight ideal cation permselective medium, wherein all three domains are of rectangular cuboid shape, as shown in FIG. 1. The left microchamber, termed "region 1", is defined in the domain $x \in [0, L_1], y \in [0, H_1], z \in [0, W_1]$, the permselective medium termed "region 2" is defined in the domain $x \in [L_1, L_1+d], y \in [0, h], z \in [0, w]$, while the right microchamber termed "region 3" is defined in the domain $x \in [L_1+d, L_1+d+L_3], y \in [0, H_3], z \in [0, W_3]$. Such a geometry realistically describes systems that have been the subject of numerous recent experimental and numerical works [6,11,12,15,21,27–32,34–37,41,42]. Additionally, this geometry can also describe a periodic array of permselective regions (e.g. nanochannel array/heterogeneous membrane) in the $y$- and or $z$-direction. The spatial coordinates have been normalized by the DL length, $\tilde{L}$. However, unlike one layer models, three layers systems can have different DL lengths on each side of the permselective medium, thus leading to a certain ambiguity [43]. So without loss of generality, we shall formulate the solution for general values of the dimensionless $L_1$ and $L_3$ while we shall remember that at least one of these values when normalized is unity (i.e. $\tilde{L}$ can be chosen arbitrarily as either $\tilde{L}_1$ or $\tilde{L}_3$).

Assuming fixed volumetric charge density, $N$, accounting for the (negative) surface charge within the nanoslot, as in classical models of permselective membranes [42,44], the space charge within all three regions $(n=1,2,3)$ can be written as follows

$$\rho_{e,n} = c_+ - c_- - N\delta_{n,2}, \qquad (6)$$

where $\delta_{n,2}$ is Kronecker's delta. The approximation $\rho_{e,n} \approx 0$ used in this study represents the LEN approximation within the microchambers along with cross-sectional electro-neutrality within the permselective medium. While both $c_+$ and $c_-$ are of order $O(1)$ in the microchamber, the case of $N \gg 1$ approximates the conditions of an ideal permselective membrane/nanochannel, i.e. $c_+ \approx N$ and $c_- \approx 0$. Obviously, this simplifying assumption does not allow the existence of intra-permselective medium concentration-polarization. This is true for membranes (e.g. Nafion) within a wide range of concentrations and nanochannels undergoing intense electric-double-layer overlap. Relaxation of this condition was recently addressed in [38].

Solution of these equations requires supplementing the appropriate boundary conditions (BC). A bulk electrolyte is defined at $x=0$ for the left microchamber (region 1) and $x=L_1+d+L_3$ for the right microchamber (region 3)

$$c(0,y,z) = c(L_1+d+L_3,y,z) = 1. \tag{7}$$

Requiring that ions do not penetrate the microchamber walls/symmetry planes ($\boldsymbol{j}_\pm \cdot \boldsymbol{n} = 0$ wherein $n$ is the coordinate normal to the surface) as well as requiring electrical insulation $\partial \phi / \partial n = 0$ at the microchamber walls/symmetry planes translates into

$$\frac{\partial c}{\partial n} = 0 \; . \tag{8}$$

This can be written explicitly as

$$c_y(x,H_i,z) = c_y(x,0,z) = c_z(x,y,W_i) = c_z(x,y,0) = 0, i=1,3 \; . \tag{9}$$

At the permselective surfaces located at $x=L_1$ and $x=L_1+d$ a simplifying assumption of uniform ionic current density along the cationic perm-selective surface (i.e. $\boldsymbol{j}_- \cdot \boldsymbol{n} = 0$) is used [16,17]

$$c_x(L_1,y,z) = \begin{cases} -i/2, & 0 \le y \le h, 0 \le z \le w \\ 0, & else \end{cases}, \tag{10}$$

$$c_x(L_1+d,y,z) = \begin{cases} -i/2, & 0 \le y \le h, 0 \le z \le w \\ 0, & else \end{cases}, \tag{11}$$

with $i$ $(=|\boldsymbol{i}|)$ being the assumed *uniform* dimensionless current density through the ion permselective boundary defined positive in the positive $x$ direction. In an ideal permselective membrane $\tilde{\boldsymbol{i}} = F\tilde{\boldsymbol{j}}_+$, or in dimensionless form $\boldsymbol{i} = \boldsymbol{j}_+$ wherein the current density has been normalized by $FDc_0/\tilde{L}$.

### C. Concentration and electric potential solutions

From Eqs. (4),(7),(9)-(11) one obtains, using the separation of variables technique for each microchamber separately, the following expressions for the 3D concentration distribution [19]

$$c_1(x,y,z) = 1 - \frac{I}{2W_1 H_1} x -$$

$$\frac{2I}{whW_1 H_1} \sum_{n,m=1}^{\infty} \frac{\sin \lambda_n^{(1)} h \sin \gamma_m^{(1)} w}{\kappa_{n,m}^{(1)} \lambda_n^{(1)} \gamma_m^{(1)} \cosh \kappa_{n,m}^{(1)} L_1} \cos \lambda_n^{(1)} y \cos \gamma_m^{(1)} z \sinh \kappa_{n,m}^{(1)} x$$

$$- \frac{I}{hW_1 H_1} \sum_{n=1}^{\infty} \frac{\sin \lambda_n^{(1)} h}{\left(\lambda_n^{(1)}\right)^2 \cosh \lambda_n^{(1)} L_1} \cos \lambda_n^{(1)} y \sinh \lambda_n^{(1)} x$$

$$- \frac{I}{wW_1 H_1} \sum_{m=1}^{\infty} \frac{\sin \gamma_m^{(1)} w}{\left(\gamma_m^{(1)}\right)^2 \cosh \gamma_m^{(1)} L_1} \cos \gamma_m^{(1)} z \sinh \gamma_m^{(1)} x$$

(12)

$$c_2(x,y,z) = N \ , \tag{13}$$

$$c_3(x,y,z) = 1 + \frac{I}{2W_3 H_3}(L_1 + d + L_3 - x) +$$

$$\frac{2I}{whW_3 H_3} \sum_{n,m=1}^{\infty} \frac{\sin \lambda_n^{(3)} h \sin \gamma_m^{(3)} w}{\kappa_{n,m}^{(3)} \lambda_n^{(3)} \gamma_m^{(3)} \cosh \kappa_{n,m}^{(3)} L_3} \cos \lambda_n^{(3)} y \cos \gamma_m^{(3)} z \sinh \kappa_{n,m}^{(3)} (L_1 + d + L_3 - x)$$

$$+ \frac{I}{hW_3 H_3} \sum_{n=1}^{\infty} \frac{\sin \lambda_n^{(3)} h}{\left(\lambda_n^{(3)}\right)^2 \cosh \lambda_n^{(3)} L_3} \cos \lambda_n^{(3)} y \sinh \lambda_n^{(3)} (L_1 + d + L_3 - x)$$

$$+ \frac{I}{wW_3 H_3} \sum_{m=1}^{\infty} \frac{\sin \gamma_m^{(3)} w}{\left(\gamma_m^{(3)}\right)^2 \cosh \gamma_m^{(3)} L_3} \cos \gamma_m^{(3)} z \sinh \gamma_m^{(3)} (L_1 + d + L_3 - x)$$

(14)

where the current is normalized by $FDc_0 \tilde{L}$

$$I = i \cdot hw \ , \tag{15}$$

and eigenvalues are defined by

$$\lambda_n^{(i)} = \frac{\pi n}{H_i}, \gamma_m^{(i)} = \frac{\pi m}{W_i}, \kappa_{nm}^{(i)} = \sqrt{\left(\lambda_n^{(i)}\right)^2 + \left(\gamma_m^{(i)}\right)^2} \ , \tag{16}$$

for regions $i = 1, 3$. The second term on the right hand side of Eqs. (12) and (14) represents the linear concentration gradient expected for a 1D problem while the remaining terms represent the geometric field-focusing [19]. Eq. (13) represents the underlying assumption of a constant counterion concentration $N$ within the ideal permselective region. It can be shown in a straight forward manner, from the Nernst-Planck relation for the cationic flux $j_+ = -\nabla c_+ - c_+ \nabla \phi \Rightarrow i = -N \, d\phi/dx$, that the potential within the ideally permselective region exhibits an Ohmic behavior

$$\phi_2(x,y) = -\frac{I}{hwN} x + \overline{\phi}_2 \ , \tag{17}$$

with $\bar{\phi}_2$ being an integration constant.

To find the three unknown constants given in Eqs. (5) and (17), we must specify additional BCs for the electric potential. The total potential drop, $V$, between $x=0$ and $x = L_1 + d + L_3$ is responsible the creation of the electric current. The BC of the electric potential at the bulk are

$$\phi = (0, y, z) = V, \quad \phi(L_1 + d + L_3, y, z) = 0 . \tag{18}$$

At the interface between the microchamber and the permselective region, the continuity of the electrochemical potential, i.e. $\mu(x, y, z) = \ln c(x, y, z) + \phi(x, y, z)$, requires that

$$\mu_1(L_1, 0, 0) = \mu_2(L_1, 0, 0) , \tag{19}$$

$$\mu_2(L_1 + d, 0, 0) = \mu_3(L_1 + d, 0, 0) . \tag{20}$$

Solving for the three BCs Eqs. (18)-(20) yields

$$\phi_1(x, y, z) = \ln[c_1(x, y, z)] + V , \tag{21}$$

$$\phi_2(x, y, z) = -\frac{I}{hwN}x + \bar{\phi}_2 , \tag{22}$$

$$\phi_3(x, y, z) = \ln[c_3(x, y, z)] , \tag{23}$$

where the constant $\bar{\phi}_2$ and I-V relations are given by

$$\bar{\phi}_2 = \frac{I}{hwN}(L_1 + d) - \ln N + 2\ln\left(1 + \frac{IL_3}{2H_3 W_3} + \bar{If}_3\right), \tag{24}$$

$$V = \frac{I}{hwN}d + 2\ln\left(\frac{1 + \dfrac{IL_3}{2H_3 W_3} + \bar{If}_3}{1 - \dfrac{IL_1}{2H_1 W_1} - \bar{If}_1}\right), \tag{25}$$

Wherein

$$\bar{f}_i = \frac{2}{whW_i H_i}\sum_{n,m=1}^{\infty}\frac{\sin\lambda_n^{(i)}h\sin\gamma_m^{(i)}w}{\kappa_{n,m}^{(i)}\lambda_n^{(i)}\gamma_m^{(i)}}\tanh\kappa_{n,m}^{(i)}L_i$$
$$+ \frac{1}{hW_i H_i}\sum_{n=1}^{\infty}\frac{\sin\lambda_n^{(i)}h}{\left(\lambda_n^{(i)}\right)^2}\tanh\lambda_n^{(i)}L_i + \frac{1}{wW_i H_i}\sum_{m=1}^{\infty}\frac{\sin\gamma_m^{(i)}w}{\left(\gamma_m^{(i)}\right)^2}\tanh\gamma_m^{(i)}L_i , \tag{26}$$

for brevity we have marked $\bar{f}_i = \bar{f}_i(L_i, H_i, h, W_i, w)$ for regions $i = 1, 3$. It is clear that the resulting I-V relation given by Eq. (25) is a function of 9 geometric parameters and the dimensionless fixed volumetric charge density $N$. A thorough analysis of the 2D and 3D behavior of the $\bar{f}_i$ functions and that of the resulting concentration has been conducted in an earlier work [19]. To simplify the subsequent analysis of 3D CP in a 3-layers system we shall present only results for the 2D case where either $w = W_1 = W_3$ or $h = H_1 = H_3$. It can be seen that the use of the former (or latter) will cancel the expressions involving the $z$ (or $y$) coordinate in the concentrations (Eqs. (12) and (14)) as well as reduce the expression in Eq. (26) to be comprised of a single term.

Before continuing, it is useful to consider the 2D case $(w = W_1 = W_3)$. In previous works [16,19], the 2D behavior of the $\bar{f}_i$ function given in Eq.(26) was thoroughly investigated. It was shown that when $h \ll H_i \ll L_i$ the function $\bar{f}_i$ $(i = 1, 3)$ reduces to

$$\bar{f}_i^{(h \ll H_i \ll L_i)} = \frac{1 - \ln(\pi h / H_i)}{\pi W_i}, \qquad (27)$$

while for the more general case of $H_i \ll L_i$ the solution was given by [19]

$$\bar{f}_i^{(H_i \ll L)} = -\frac{H_i}{\pi^2 h W_i} \operatorname{Re}\left[ jLi_2\left(e^{j\pi h/H_i}\right) \right], \qquad (28)$$

where Re is the real part, $j$ is the imaginary unit and $Li_2(\theta)$ is the polylogarthim of order 2 and argument $\theta$. The two key points that are apparent in Eqs. (27) and (28) are firstly that these solutions are functions of the degree of heterogeneity of the system $h/H_i$ with 1 being a homogeneous system and 0, a completely heterogeneous system. Secondly, as this ratio, $h/H_i$, goes to zero, these functions (Eqs. (26)-(28)) become singular and approach infinity. As we have previously shown [19] the heterogeneity in the third dimension only adds to that in already existing in the 2D in plane problem and further increases the effects of geometric field focusing.

### III. Numerical simulations

To verify our results we solved the PNP equations given by Eq. (1)-(3) using the finite elements program Comsol$^{TM}$ for the 2D geometry described in FIG. 2. Unlike the above theoretical model, the numerical model accounts for non-electroneutral effects, non-ideality of the permselective region. Thus, we solved the case of $\delta = 10^{-3}$. Comparison of theoretical and numerical results will be conducted in the next section. For additional information on implementation of electrodiffusive simulations in Comsol see supplemental information of Ref. [15].

## IV. Results

### A. Concentration and electric potential

FIG. 3 shows a 2D plot of the concentration distribution for under-limiting current conditions. Equi-concentration contours qualitatively illustrate the radial concentration gradients towards the microchamber-membrane interface while a 1D linear concentration gradient is obtained further away from the interface at a radial distance $r \sim H_1$. As was previously discussed [11], in the limit of infinitely large reservoirs, the 2D concentration profile has a logarithmic dependency on the radial distance. Thus, increased heterogeneity $(h/H_{1,3} \to 0)$ results in an effectively shorter depletion length when compared to the 1D case of a linear distribution. FIG. 4 shows a comparison between our simplified LEN theoretical model and simulation of the fully coupled PNP equations, which accounts for the creation of the SCL, for the concentration and electric potential profiles along $y=0$. An excellent agreement is obtained, thus, confirming the validity of our approximation for small voltages and Debye layers. It is clearly shown that the concentration gradients within the right microchamber are larger than at the left so as to compensate for its smaller height in order to sustain continuity of current.

### B. Current-voltage curves and overall system conductance

FIG. 5 plots the I-V curves for symmetric microchambers in a 2D system where the height of either the microchamber (FIG. 5a) or the height of the permselective region are varied (FIG. 5b). It can be seen that as the microchamber height increases, so too does the conductance, i.e. slope of the I-V curve at the Ohmic regime, and the limiting current. This can be expected as the overall system resistance decreases with increasing microchamber height. In contrast, the average current density $\bar{i} = I/HW$ shows a reversal with increasing system heterogeneity, i.e. it increases as the microchamber height decreases (inset of FIG. 5a). A similar trend is shown for the current density $i = I/hw$ when the microchamber height is kept constant while the permselective region height is varied, indicating that the current density increases with increased heterogeneity (inset of FIG. 5b).

Based on the above analysis, it is evident that in the low voltage/Ohmic regime the slope of the current is dependent on the microchannel geometry. This indicates that the conductance of the system may no longer be solely dependent on the permselective region geometry as commonly assumed in microchannel/nanochannel systems [11,12,30,31,36].

In the Ohmic region, for the case of small currents $(I \ll 1)$, Eq. (25) is expanded

$$V = I\left(\frac{d}{hwN} + \frac{L_1}{H_1 W_1} + \frac{L_3}{H_3 W_3} + 2\bar{f_1} + 2\bar{f_3}\right), \qquad (29)$$

hence, the overall conductance (normalized by $DF^2 c_0 \tilde{L}/RT$) of the 3-layers system is given by

$$\sigma_{N\gg 1} = \frac{I}{V} = \left( \frac{d}{hwN} + \frac{L_1}{H_1 W_1} + \frac{L_3}{H_3 W_3} + 2\overline{f_1} + 2\overline{f_3} \right)^{-1}, \tag{30}$$

Rewriting the conductance in dimensional form yields

$$\tilde{\sigma}_{N\gg 1} = \frac{\tilde{I}}{\tilde{V}} = \frac{DF^2 c_0}{RT} \left( \frac{\tilde{d}}{\tilde{h}\tilde{w}\tilde{N}/c_0} + \frac{\tilde{L}_1}{\tilde{H}_1 \tilde{W}_1} + \frac{\tilde{L}_3}{\tilde{H}_3 \tilde{W}_3} + 2\tilde{\overline{f_1}} + 2\tilde{\overline{f_3}} \right)^{-1}, \tag{31}$$

wherein from Eq. (26) $\overline{f}_i = \tilde{\overline{f}}_i \cdot \tilde{L}$. Eq. (31) is valid in the limit of $\tilde{N} \gg c_0 (N \gg 1)$. In the case of a straight nanochannel, the surface charge density, $\tilde{\sigma}_s$, and $\tilde{N}$ are related by the following relation

$$\tilde{N} = \frac{\alpha |\tilde{\sigma}_s|}{zF} \left( \frac{1}{\tilde{h}} + \frac{1}{\tilde{w}} \right), \tag{32}$$

wherein $\alpha = 1$ when the permselective medium bottom and side surfaces ($x \in [L_1, L_1 + d]$, $y = 0$ and $z = 0$) are symmetry planes, and $\alpha = 2$ when these are physical surfaces that are charged.

Alternatively, for the case of charged porous medium [45] of porosity $\varepsilon_p$, internal pore surface area/volume $\tilde{a}_p$ and pore surface charge/area $\tilde{\sigma}_s$, the surface charge per pore volume is

$$\tilde{N} = \frac{\tilde{a}_p |\tilde{\sigma}_s|}{zF\varepsilon_p}. \tag{33}$$

The physical meaning of each term in Eq. (31) becomes more evident for the 1D case where $H_{1,3} = h$ and $W_{1,3} = w$, hence $\overline{f}_{1,3} = 0$. Then, it can immediately be seen that the first term is the resistance of the permselective region, while the second and third are the resistors of the microchambers. The fourth and fifth term are resistances that can be attributed to the geometric field focusing effects occurring at the two microchamber-permselective medium interfaces. When the first term is significantly larger than the remaining terms, i.e., resistance of the permselective region dominates, one expectedly finds that the conductance is independent of the microchamber geometry (i.e. all the curves in FIG. 5 would collapse onto the same curve). However this is not necessarily true when the microchambers resistance is approaching that of the permselective medium.

Yossifon and coworkers [11,12] derived an expression for the conduction (per unit width) of nanoslot-dominated system, valid across the entire range of concentrations, using the well-known Donnan potential equilibrium relations [39,42]

$$\tilde{\sigma} = 2\frac{DF^2}{RT}\frac{\tilde{h}\tilde{w}}{\tilde{d}}\sqrt{c_0^2 + \left(\frac{\tilde{N}}{2}\right)^2} \quad . \tag{34}$$

In the limit of an electrolyte with low concentration/ideal permselectivity, i.e. $\tilde{N} \gg c_0$, the conductance in Eq. (34) is identical to that of Eq. (31) when the microchamber and field-focusing resistances are neglected. In the other extreme limit of a highly concentrated electrolyte/vanishing permselectivity, $\tilde{N} \ll c_0$, Eq. (34) yields a linear dependency of the conductance on the concentration

$$\tilde{\sigma} = 2\frac{DF^2}{RT}\frac{\tilde{h}\tilde{w}}{\tilde{d}}c_0 \quad . \tag{35}$$

For the limit of vanishing permselectivty of region 2 the normalized conductance is given by (see Appendix )

$$\sigma_{N \ll 1} = 2\left(\frac{d}{hw} + \frac{L_1}{H_1 W_1} + \frac{L_3}{H_3 W_3} + 2\overline{f_1} + 2\overline{f_3}\right)^{-1} \quad , \tag{36}$$

and in dimensional form

$$\tilde{\sigma}_{N \ll 1} = 2\frac{DF^2 c_0}{RT}\left(\frac{\tilde{d}}{\tilde{h}\tilde{w}} + \frac{\tilde{L}_1}{\tilde{H}_1 \tilde{W}_1} + \frac{\tilde{L}_3}{\tilde{H}_3 \tilde{W}_3} + 2\tilde{\overline{f_1}} + 2\tilde{\overline{f_3}}\right)^{-1} \tag{37}$$

For the case where the permselective region resistance dominates Eq. (37) reduces to Eq.(35) above. In FIG. 6 we compare the conductance (per unit width i.e 2D system) given by Eqs. (31), (34), and (37) to the numerical results calculated for a symmetric geometry. The simulations were conducted in dimensional form with the same BCs givens in FIG. 2 for bulk concentration, $c_0$, varying from $10^{-3}$ to $10^3 \left[mol/m^3\right]$ and fixed $\tilde{N} = 0.76 \left[mol/m^3\right]$ which is based on the values taken from Ref. [12]. Eq.(31), which accounts for the microchamber and field focusing resistances, captures an interesting phenomenon -the divergence of the conductance from a constant value at low concentrations/ideal selectivity. This is confirmed numerically. In common micro-nanochannel devices, the small size of the permselective height $\tilde{h}$ and relatively large length $\tilde{d}$ ensure the dominance of the permselective resistance [27,30,31] However, depending on the geometry of the system and at low enough concentrations, the microchamber and field focusing resistances can be comparable to that of the nanochannel, as was observed in Yossifon et al. [28] where the conductance in the Ohmic region was weakly dependent on the microchamber height (also evident in the inset Fig. 5b of Ref [19]).

### C. Limiting current and current rectification

Limiting current occurs when the concentration at the interface of the microchambers and permselective region drops to zero, which from Eqs.(12) and (14) we then obtain

$$I_{\lim,1} = \left(\frac{L_1}{2H_1 W_1} + \bar{f}_1\right)^{-1}, \tag{38}$$

$$I_{\lim,3} = -\left(\frac{L_3}{2H_3 W_3} + \bar{f}_3\right)^{-1}. \tag{39}$$

corresponding to $c_1(L_1,0,0) = 0$ and $c_3(L_1+d,0,0) = 0$, respectively. It is clearly seen the limiting current increases with increasing microchamber height and width. Also, it is clear that the limiting current at each of the microchambers is oblivious to the geometry of the counter microchamber.

That the limiting current is solely determined by the geometry of the microchamber undergoing depletion, suggesting that asymmetric microchamber geometries will result in different limiting currents. Hence, one should expect current rectification under opposite polarization of the externally applied fields. The rectification factor is defined as

$$R = \left|\frac{I_{V>0}}{I_{V<0}}\right|, \tag{40}$$

where a positive voltage corresponds to depletion in region 1 and negative voltages correspond to depletion in region 3. Current rectification is expected to occur whenever a geometrical asymmetry is introduced into the system. Here it is demonstrated for the case where only the right microchamber height (region 3) is varied. It is seen in FIG. 7 that when the anodic side is at the left microchamber with a fixed depth (region 1), the limiting currents eventually collapse onto a single curve. In contrast, when the depth of the anodic side of the microchamber is varied (region 3), we obtain different limiting currents, indicating current rectification. That rectification occurs also for low voltages (in the Ohmic region) is confirmed both by our theoretical model as well as by simulations.

### V. Conclusions

In this work we have studied the effects of geometry on both the 2D and 3D concentration and electric potential distributions for a 3-layers system undergoing CP due to the application of an external electric potential or current. An analytical solution of the electro-diffusive problem, under the LEN approximation, was obtained using the separation of variables technique for the two opposing asymmetric microchambers. Assuming an ideal permselective medium allows for the analytic calculation of the 3D concentration and electric potential distributions as well as a current-voltage relation.

The solutions for the concentration distributions include the standard 1D linear terms as well as additional terms that account for the 2D and 3D geometric field focusing and are functions of the heterogeneity of the system, i.e $h/H_i \neq 1$ and/or $w/W_i \neq 1$. It is shown that as these ratios decrease below 1, the effects of heterogeneity increase, indicating the increase of the geometric field focusing effects and with this also the intensification of the current density (FIG. 5). It can be deduced from our analysis that any system is inherently 3D unless homogeneity exists in a certain direction, reducing the system to be either 2D or 1D. It is shown that any asymmetry in the microchamber geometries will result in current rectification at any voltage, even within the Ohmic regime, due to CP (FIG. 7). Moreover, the overall conductance of the system is derived from the Ohmic response of the resistance in all three regions, i.e. the depleted and enriched diffuse layers as well as the permselective medium, as well as the geometrical field-focusing effects. It is demonstrated that for non-negligible microchamber resistances the conductance does not exhibit the expected saturation at low concentrations but rather shows a continuous decrease (FIG. 6).

The resulting analytical relation for the current-voltage will facilitate a more direct comparison between theory and experiments as now the voltage drop is across the entire realistic 3D and 3-layer system. The theoretical framework provided in this work for a 3D and 3-layers systems can be expanded upon to account for additional effects, e.g. asymmetric bulk concentrations [21,22], non-straight nanochannel/nanopore geometries [32], or even asymmetric entrances of a straight permselective medium [14].

**Acknowledgments**
We wish to acknowledge J.Schiffbauer and A.Boymelgreen for their input. This work was supported by ISF Grant No. 1078/10. We thank the Technion RBNI (Russell Berrie Nanotechnology Institute) and the Technion GWRI (Grand Water Research Institute) for their financial support.

### Appendix : Derivation of conductance at the limit of vanishing permselectivity

For the case of vanishing permselectivity (sufficiently high concentrations, i.e. $N \ll 1$) electric Debye layer overlap does not occur (i.e. $\delta/h \ll 1$), and region 2 as described in FIG. 1 is no longer permselective. Hence, a uniform concentration $(c = 1)$ exists throughout the system. Once more we assume a symmetric and binary electrolyte as in Section II.A. Thus, the governing equation for the potential throughout the 3 layers system is simply the Laplace equation

$$\nabla^2 \phi = 0 \ . \tag{41}$$

The BCs are (FIG. 1) the applied potentials at the microchamber-bulk interfaces

$$\phi(0, y, z) = V, \quad \phi(L_1 + d + L_3, y, z) = 0 \ . \tag{42}$$

The electrical insulation condition are

$$\phi_y(x, H_i, z) = \phi_y(x, 0, z) = \phi_z(x, y, W_i) = \phi_z(x, y, 0) = 0 \quad i = 1, 3, \quad (43)$$

and uniform ionic current density at the interface of the microchamber and permselective region

$$\phi_x(L_1, y, z) = \begin{cases} -i/2, & 0 \leq y \leq h, 0 \leq z \leq w \\ 0, & else \end{cases}, \quad (44)$$

$$\phi_x(L_1 + d, y, z) = \begin{cases} -i/2, & 0 \leq y \leq h, 0 \leq z \leq w \\ 0, & else \end{cases}. \quad (45)$$

It should be noted that now the current density has a symmetric Ohmic contribution from both the cations and anions, i.e. $\tilde{i} = F(\tilde{j}_+ - \tilde{j}_-) = -2\frac{DF^2}{RT}\tilde{c}\tilde{\nabla}\tilde{\phi}$ or in dimensionless form $i = -2c\nabla\phi = -2\nabla\phi$.

From observation once concludes that the boundary value problem for the electric potential, Eqs. (41)-(45), is identical to that for the ionic concentration, Eqs.(4), (7) and (9)-(11), within the microchambers (i.e. regions 1 and 3) except at the bulk interfaces (i.e. Eq.(42) versus Eq.(7)). This immediately suggests the following solution based on that obtained for the concentration using the separation of variable technique

$$\phi_1(x, y, z) = V - \frac{I}{2W_1 H_1} x -$$
$$\frac{2I}{whW_1 H_1} \sum_{n,m=1}^{\infty} \frac{\sin \lambda_n^{(1)} h \sin \gamma_m^{(1)} w}{\kappa_{n,m}^{(1)} \lambda_n^{(1)} \gamma_m^{(1)} \cosh \kappa_{n,m}^{(1)} L_1} \cos \lambda_n^{(1)} y \cos \gamma_m^{(1)} z \sinh \kappa_{n,m}^{(1)} x, \quad (46)$$

$$-\frac{I}{hW_1 H_1} \sum_{n=1}^{\infty} \frac{\sin \lambda_n^{(1)} h}{\left(\lambda_n^{(1)}\right)^2 \cosh \lambda_n^{(1)} L_1} \cos \lambda_n^{(1)} y \sinh \lambda_n^{(1)} x$$
$$-\frac{I}{wW_1 H_1} \sum_{m=1}^{\infty} \frac{\sin \gamma_m^{(1)} w}{\left(\gamma_m^{(1)}\right)^2 \cosh \gamma_m^{(1)} L_1} \cos \gamma_m^{(1)} z \sinh \gamma_m^{(1)} x$$

$$\phi_2(x, y, z) = -\frac{I}{2hw} x + \overline{\phi}_2, \quad (47)$$

$$\phi_3(x,y,z) = \frac{I}{2W_3 H_3}(L_1 + d + L_3 - x) +$$

$$\frac{2I}{whW_3 H_3} \sum_{n,m=1}^{\infty} \frac{\sin \lambda_n^{(3)} h \sin \gamma_m^{(3)} w}{\kappa_{n,m}^{(3)} \lambda_n^{(3)} \gamma_m^{(3)} \cosh \kappa_{n,m}^{(3)} L_3} \cos \lambda_n^{(3)} y \cos \gamma_m^{(3)} z \sinh \kappa_{n,m}^{(3)} (L_1 + d + L_3 - x)$$

$$+ \frac{I}{hW_3 H_3} \sum_{n=1}^{\infty} \frac{\sin \lambda_n^{(3)} h}{\left(\lambda_n^{(3)}\right)^2 \cosh \lambda_n^{(3)} L_3} \cos \lambda_n^{(3)} y \sinh \lambda_n^{(3)} (L_1 + d + L_3 - x) \quad , \quad (48)$$

$$+ \frac{I}{wW_3 H_3} \sum_{m=1}^{\infty} \frac{\sin \gamma_m^{(3)} w}{\left(\gamma_m^{(3)}\right)^2 \cosh \gamma_m^{(3)} L_3} \cos \gamma_m^{(3)} z \sinh \gamma_m^{(3)} (L_1 + d + L_3 - x)$$

where the eigenvalues are given by the previous relation (Eq.(16)).

Requiring continuity of the electric potential, instead of the electrochemical potential as before, at the interfaces of the microchambers and the permselective region (i.e. $\phi_1(L_1, 0, 0) = \phi_2(L_1, 0, 0)$ and $\phi_2(L_1 + d, 0, 0) = \phi_3(L_1 + d, 0, 0)$) one obtains

$$\bar{\phi}_2 = \frac{I}{2hw}(L_1 + d) + \frac{IL_3}{2H_3 W_3} + I\bar{f}_3 \quad , \quad (49)$$

$$V_{N \ll 1} = \frac{I}{2}\left(\frac{d}{hw} + \frac{L_1}{H_1 W_1} + \frac{L_3}{H_3 W_3} + 2\bar{f}_1 + 2\bar{f}_3\right) . \quad (50)$$

$$\sigma_{N \ll 1} = 2\left(\frac{d}{hw} + \frac{L_1}{H_1 W_1} + \frac{L_3}{H_3 W_3} + 2\bar{f}_1 + 2\bar{f}_3\right)^{-1} , \quad (51)$$

where it is once more clear that an equivalent circuit is comprised of three geometry dependent resistors and two field focusing resistors. It is noted that the $\bar{f}_i$ functions are reminiscent of the access and convergence resistance terms calculated for a single and isolated nanopore [46,47]. In these works it was shown that resistance resulting from field focusing increases as the nanopore radius decreases. Similarly, the $\bar{f}_i$ functions increases with increasing heterogeneity, i.e. divergence of $h/H_{1,3}$ and $w/W_{1,3}$ from 1 [19].

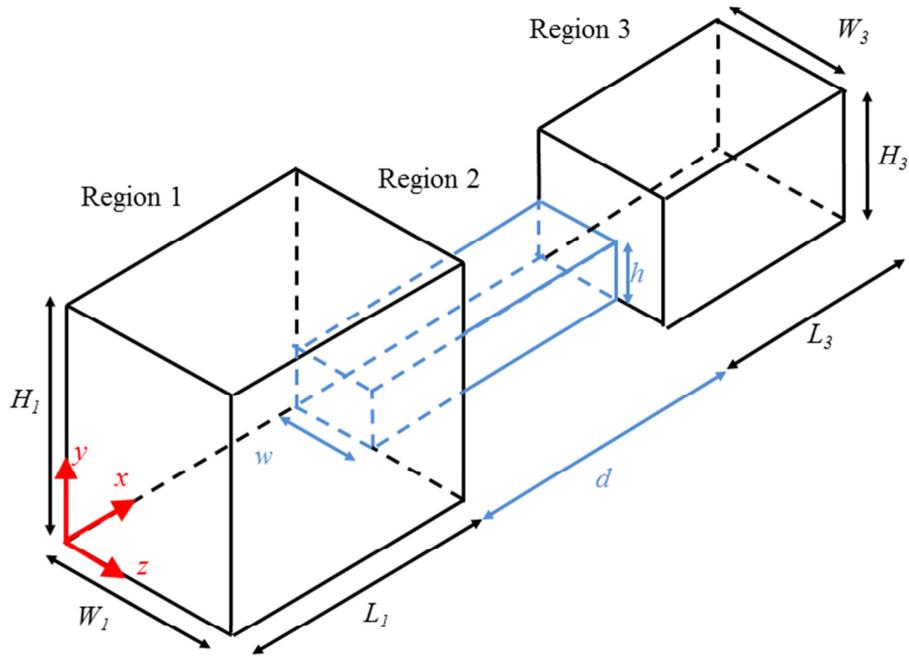

**FIG. 1. (Color online) Schematics describing the 3D geometry of the three layer system consisting of a straight permselective medium connecting two opposite asymmetric microchambers.**

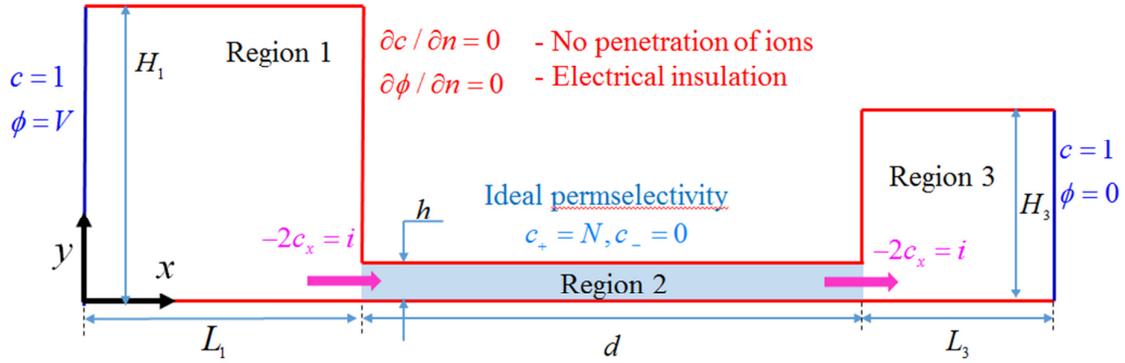

**FIG. 2.** (Color online) Schematics describing the 2D geometry (i.e. $w = W_1 = W_3$) of the three layer system consisting of a straight permselective medium connecting two opposite asymmetric microchambers. The electrodiffusive boundary conditions have also been added for clarity. In most of the subsequent analysis, the 2D geometry shall be used for demonstration purposes.

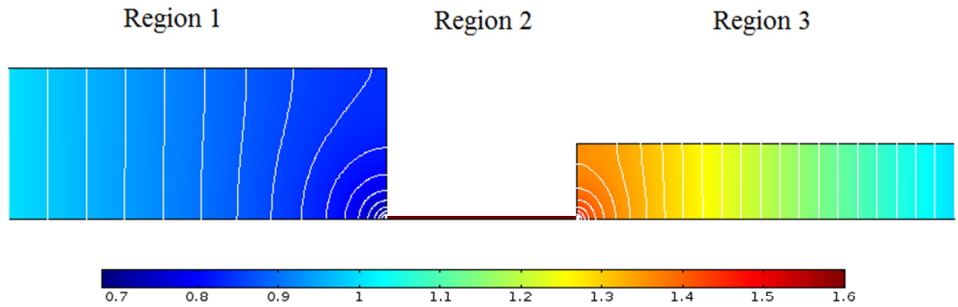

**FIG. 3. (Color online)** The 2D (i.e. $w = W_1 = W_3$) concentration distribution plot obtained from numerical simulation for the following geometry: $L_1 = L_3 = 1$, $d = 0.5$, $H_1 = 0.4$, $H_3 = 0.2$, $h = 0.01$, $N = 25$, $\delta = 10^{-3}$ at $V = 2$ exhibiting concentration polarization, i.e. depletion and enrichment within regions 1 and 3, respectively.

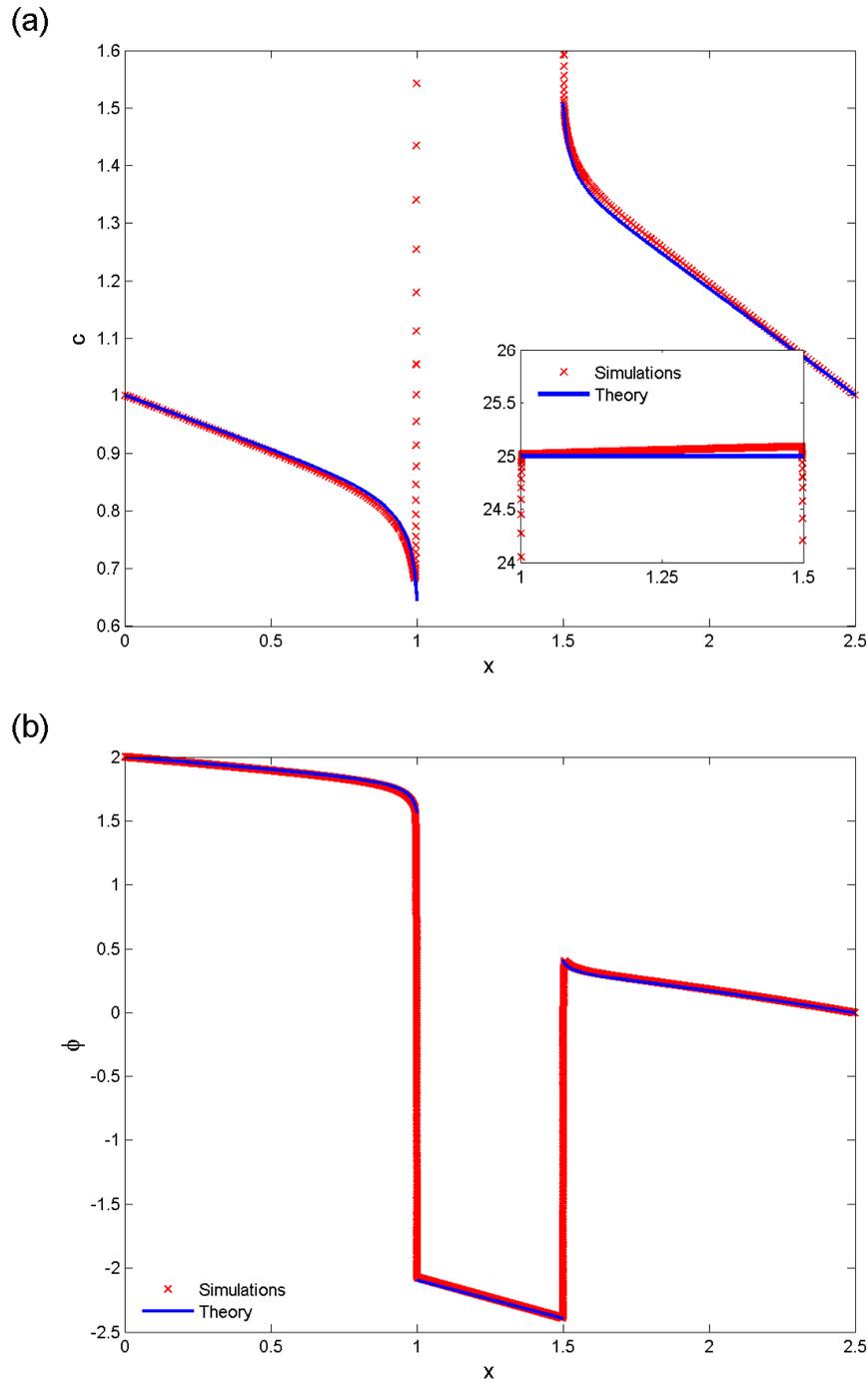

FIG. 4. (Color online) (a) The concentration and (b) electric potential profiles along the bottom surface $y=0$ are compared between theory and numerical simulations $(\delta=10^{-3})$ for the asymmetric microchamber geometry depicted in FIG. 3. The inset of part (a) shows the concentration of the permselective region which is substantially higher than the reservoirs with a fixed counterion concentration for the ideal case (theory) and a slight concentration gradient for the numerical simulation.

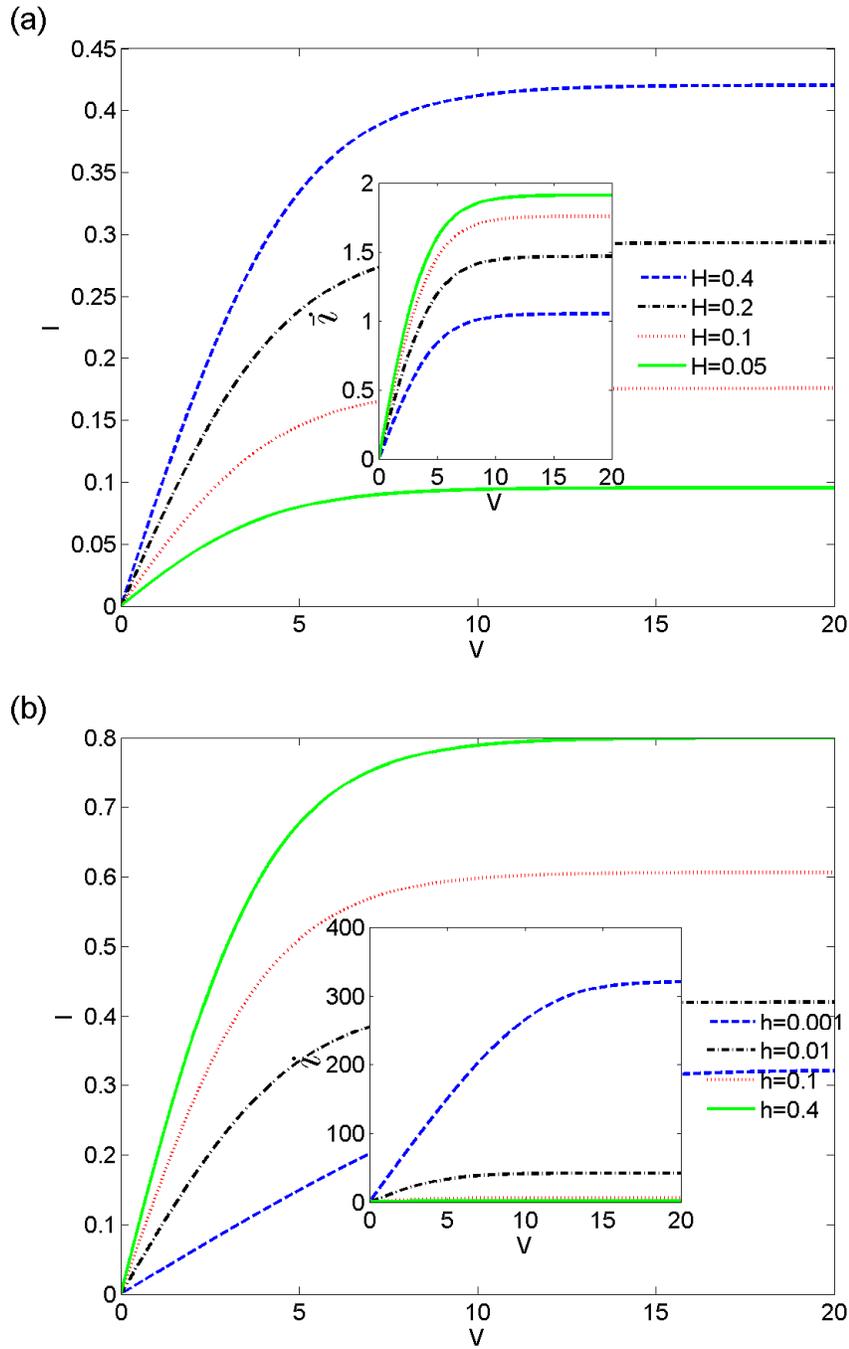

FIG. 5. (Color Online) I-V curves for 2D device (i.e. $w = W_1 = W_3 = 1$) with symmetric microchambers, $L_1 = L_3 = 1$, $H_1 = H_2 = H$. The length of the nanoslot, $d = 0.5$ and average volumetric concentration $N = 25$ are kept constant. (a) $H$ is varied while the permselective region height is kept constant $h = 0.01$. The inset shows that the average current density $\bar{i} = I/HW$. (b) The microchamber height is kept constant at $H = 0.4$ and $h$ is varied. The inset shows the current density $i = I/hw$.

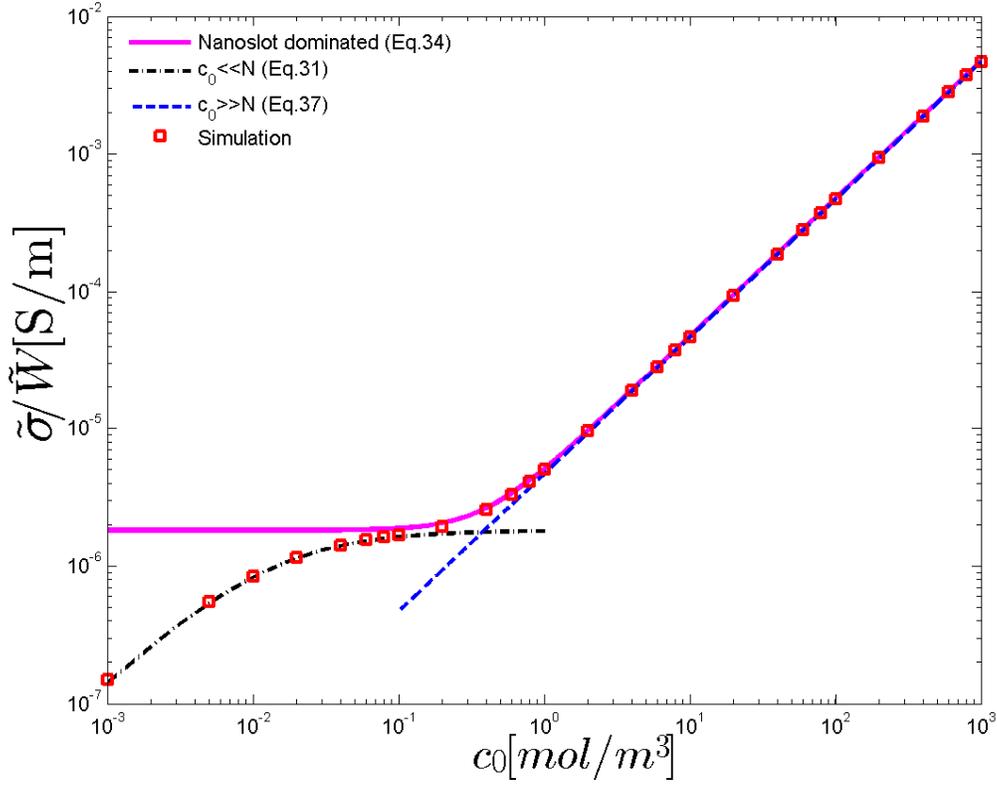

**FIG. 6. (Color online) Comparison of various conductance (per unit width) asymptotic models versus the fully coupled numerical simulation for a symmetric microchamber geometry of a 2D device (i.e $\tilde{w} = \tilde{W}_1 = \tilde{W}_3 = \tilde{W}$ ) $\tilde{L}_1 = \tilde{L}_3 = \tilde{d}/3 = 100[\mu m], \tilde{H}_1 = \tilde{H}_3 = 10[\mu m], \tilde{h} = 190[nm], \tilde{N} = 0.76[mol/m^3]$. A clear divergence from saturation of the conductance at low concentrations is obtained when the microchambers resistance is not negligible.**

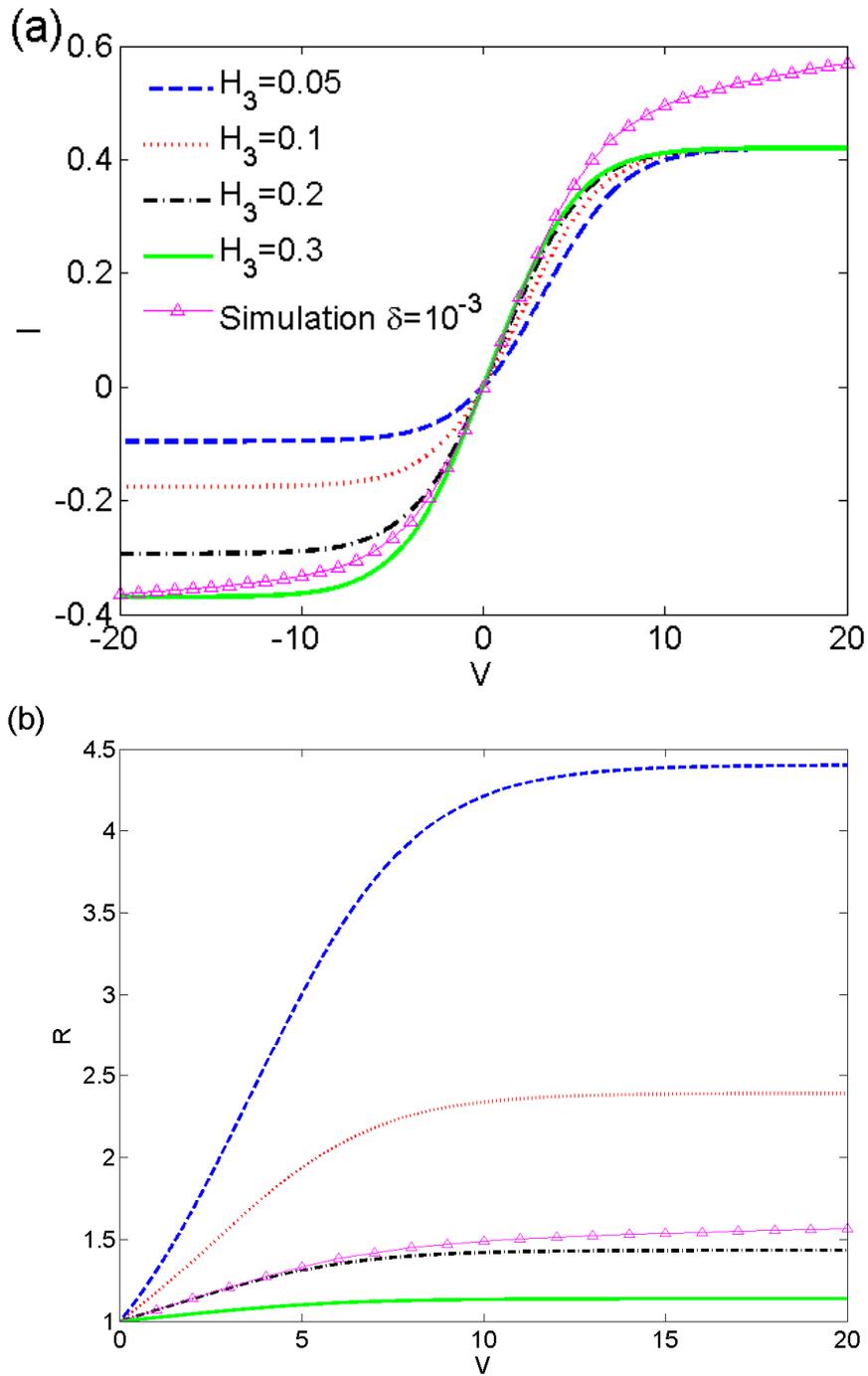

FIG. 7. (Color online) (a) I-V and (b) rectification curves for a 2D device (i.e. $w = W_1 = W_3 = 1$) consisting of microchambers with $L_1 = L_3 = 1$, $h = 0.01$, $d = 0.5$, $H_1 = 0.4$, $N = 25$ and varying $H_3$.